\documentclass[12pt]{elsart}
\usepackage{epsfig}

\newcommand{\nn}{\nonumber}
\newcommand{\raw}{\rightarrow}

\newcommand{\bi}{\begin{itemize}}
\newcommand{\ei}{\end{itemize}}
\newcommand{\be}{\begin{equation}}
\newcommand{\ee}{\end{equation}}
\newcommand{\bea}{\begin{eqnarray}}
\newcommand{\eea}{\end{eqnarray}}

\def      \nue{\ensuremath{\nu_{e}}\ }
\def      \nuebar{\ensuremath{\overline{\nu}_{e}}\ }

\def      \numu{\ensuremath{\nu_{\mu}}\ }

\def      \simge{\mathrel{%
   \rlap{\raise 0.511ex \hbox{$>$}}{\lower 0.511ex \hbox{$\sim$}}}}
\def      \simle{\mathrel{
   \rlap{\raise 0.511ex \hbox{$<$}}{\lower 0.511ex \hbox{$\sim$}}}}
\def      \BB{$\beta$-Beam\ }
\def      \SB{Super-Beam\ }

\begin{document}
%
\begin{frontmatter}

\title{Appearance and disappearance signals at a $\beta$-Beam and a 
Super-Beam facility}

\author{A. Donini},
\author{E. Fern\'andez-Mart\'{\i}nez} and
\author{S. Rigolin}
\address{Instituto de Fisica Teorica and Departamento de F\'{\i}sica Te\'{o}rica, 
Universidad Autonoma de Madrid, E-28049, Madrid, Spain}

\begin{abstract}
In this letter we present the study of the eightfold degeneracy in the 
$(\theta_{13},\delta)$ measurement including both appearance and disappearance channels. 
We analyse, for definiteness, the case of a standard low-$\gamma$ $\beta$-Beam 
and a 4 MWatt SPL Super-Beam facility, both aiming at a UNO-like Mton water \v Cerenkov
detector located at the Fr\'{e}jus laboratory, $L=130$ km. In the $\beta$-Beam case, 
the \nue disappearance channel does not improve the $(\theta_{13},\delta)$ measurement 
when a realistic (i.e. $\ge$ 2\%) systematic error is included. In the Super-Beam case, 
the \numu disappearance channel could, instead, be quite useful in reducing the impact 
of the eightfold degeneracy in the $(\theta_{13},\delta)$ measurement, especially once 
the error on the atmospheric mass difference is fully taken into account in the fit. 
\end{abstract}

\begin{keyword}
Neutrino Oscillation \sep Super-Beam \sep $\beta$ Beam
\PACS: 14.60.Pq \sep 14.60.Lm
\end{keyword}

\end{frontmatter}

%
%
\newpage
%
%
\section{Introduction}
\label{introduction}

After more than 30 years of successful neutrino oscillation experiments \cite{exp} two 
parameters still remain undetermined in the three-family Pontecorvo-Maki-Nakagawa-Sakata 
\cite{neutrino_osc} mixing matrix: the mixing angle $\theta_{13}$, for which only a upper 
limit has been set \cite{chooz}, and the CP-violating phase $\delta$ that is still completely 
unknown. The full understanding of the leptonic mixing matrix constitutes, together with the 
discrimination of the Dirac/Majorana character and the measure of the neutrino absolute mass 
scale, the main neutrino-physics goal for the next decade(s). 

It is well known that the best way to simultaneously measure $(\theta_{13},\delta)$ 
is the (golden) $\nue \!\!\raw \numu$ appearance channel \cite{Cervera:2000kp} 
(and its T and CP conjugate ones). Unfortunately this measure is, in general, severely 
affected by the presence of degeneracies. When a measurement of the two unknown parameters 
is performed using a beam able to produce both neutrinos and antineutrinos, the following 
four systems of equations must be solved:
\be
\left \{ 
\begin{array}{lll}
N_{l^+} (\bar \theta_{13},\bar \delta; \bar s_{atm}, \bar s_{oct}) &=& 
N_{l^+} (\theta_{13},\bar \delta; \pm \bar s_{atm}, \pm \bar s_{oct}) \, , \\
N_{l^-} (\bar \theta_{13},\bar \delta; \bar s_{atm}, \bar s_{oct}) &=& 
N_{l^-} (\theta_{13},\bar \delta; \pm \bar s_{atm}, \pm \bar s_{oct}) \, , 
\end{array}
\right .
\ee
where $s_{atm} = sign(\Delta m^2_{atm})$ and $s_{oct} = sign(\tan 2 \theta_{23})$ 
are two discrete unknowns, the sign of the atmospheric mass difference and the
$\theta_{23}$-octant. The r.h.s. of this equation implies that four different 
models (each of them with a definite $(s_{atm},s_{oct})$ choice) must be used 
to fit the data on the l.h.s. The eight solutions form what is known as the 
{\it eightfold degeneracy} \cite{Burguet-Castell:2001ez}-\cite{Barger:2001yr}.
Various methods have been considered to get rid of degeneracies (using spectral 
analysis \cite{Burguet-Castell:2001ez}, combination of experiments \cite{complementSB} 
and/or different channels \cite{Donini:2002rm}). In principle, the eightfold 
degeneracy can be completely solved if a sufficient number of independent informations 
is added. At the cost, of course, of increasing the number of detectors and/or beams 
and consequently the budget needs. 

In this letter we try to understand if the effect of degeneracies can be reduced 
using informations from both the appearance and the disappearance channels at a 
given experiment. We consider, as reference, the proposal for two CERN-based 
facilities, the standard\footnote{Other \BB proposals with different choices
of the boosting factor can be found in \cite{Volpe:2003fi}-\cite{Terranova:2004hu}.}
low-$\gamma$ \BB \cite{Zucchelli:sa} and the \SB based on the 4 MWatt SPL 2.2 GeV 
proton driver \cite{Gomez-Cadenas:2001eu}. Both beams are directed from CERN toward 
the underground Fr\'ejus laboratory, where it has been proposed to locate a 1 Mton 
UNO-like \cite{Jung:1999jq} water \v Cerenkov detector with a 440 kton fiducial mass. 
The considered baseline is $L=130$ km. To be at the first peak in the leading 
oscillation probability term, the average neutrino energy for both beams has been 
chosen of the order of a few hundreds MeV. Of course, many other similar setups 
could be considered, instead the ``standard'' ones adopted in this letter. Anyway 
our considerations are quite general and will hold for any comparable low-$\gamma$ 
\BB and \SB setup. 

Needless to say that a similar analysis can be performed in any experiment 
where disappearance and appearance channels are simultaneously available. 
At the Neutrino Factory, for example (see \cite{Apollonio:2002en,Blondel:2000gj}), 
the $\bar\nu_\mu$ disappearance channel can be certainly used together with the 
appearance channel $\nu_e \to \nu_\mu$, whereas the $\nu_e$ disappearance channel is 
extremely difficult to exploit (due to the need to measure the electron charge to 
distinguish $\nu_e \to \nu_e$ from $\bar \nu_\mu \to \bar \nu_e$). 

The eightfold degeneracy for these two facilities has been comprehensively studied 
in \cite{Donini:2004hu} and we refer to that paper for all the technical details 
regarding the used cross-sections, efficiencies and backgrounds. The results of 
\cite{Donini:2004hu} show that to run the two facilities simultaneously does not 
help in solving the degeneracies, mainly because the two beams, running on the 
same baseline and with approximately the same energy, are not complementary at all.
The only effect is to increase the statistics by roughly a factor two and to reduce 
some of the systematics, but leaving practically unaffected the main systematic error 
that it's due to the definition of the fiducial volume of a Mton water detector. In 
this sense no real {\it synergy} is achieved adding these two experiments (if not 
for using the same detector, thus halving the corresponding costs). For this reason, 
in the following we will analyse the performance (in the appearance and disappearance 
channels) of the two facilities separately.\footnote{We will not consider here the 
possibility of using the \BB or \SB facility for other measures beyond appearance 
and disappearance oscillation ones. The interested reader can find a detailed 
description of these other possible measurements in, for example, 
\cite{Zucchelli:sa,Volpe:2003fi,Bouchez:2003fy}.} 

%
%
\section{$\beta$-Beam Appearance and Disappearance Channels}
%
%

The considered \BB setup consists of a $\bar \nu_e$-beam produced by the decay of 
$^6$He ions boosted at $\gamma = 60$ and of a $\nu_e$-beam produced in the decay of 
$^{18}$Ne ions boosted at $\gamma = 100$. The $\gamma$-ratio has been chosen to store 
both ions simultaneously into the decay ring. A flux of $2.9 \times 10^{18}$ $^6$He 
decays/year and $1.1 \times 10^{18}$ $^{18}$Ne decays/year, as discussed in 
\cite{Bouchez:2003fy}, will be assumed. The average neutrino energies of the 
$\nu_e,\bar \nu_e$ beams corresponding to this configuration are 0.37 GeV and 0.23 GeV, 
respectively. Although the boosting factor has been chosen to maximize the oscillation 
probability at $L = 130$ km, a severe drawback of this option is the impossibility to 
use energy resolution, due to nuclear effects. 
\begin{table}[hbtp]
\begin{center}
\begin{tabular}{|c|c|c|c|c|c|} \hline
 & No Osc. & $\theta_{13} = 8^\circ; +$ & $\theta_{13} = 8^\circ; -$ & 
             $\theta_{13} = 2^\circ; +$ & $\theta_{13} = 2^\circ; -$\\ \hline \hline
$N_{e^-}$ & 133205 &  89426 & 89742 & 93837 & 93865 \\ \hline
$N_{e^+}$ & 19557 & 12180 & 12158 & 13000 & 12999 \\ \hline
\hline
\end{tabular}
\end{center}
\caption{\it 
Disappearance event rates for a $10$ years run at the considered \BB with a $440$ kton 
detector at $L = 130$ km, for different values of $\theta_{13}$ and of the sign of the 
atmospheric mass difference, $s_{atm}$. Appearance event rates have been quoted in 
ref.~\cite{Donini:2004hu}. }
\label{tab:betabeam}
\end{table}

\begin{figure}[t!]
\vspace{-0.5cm}
\begin{center}
\begin{tabular}{cc}
\hspace{-1.0cm} \epsfxsize8.25cm\epsffile{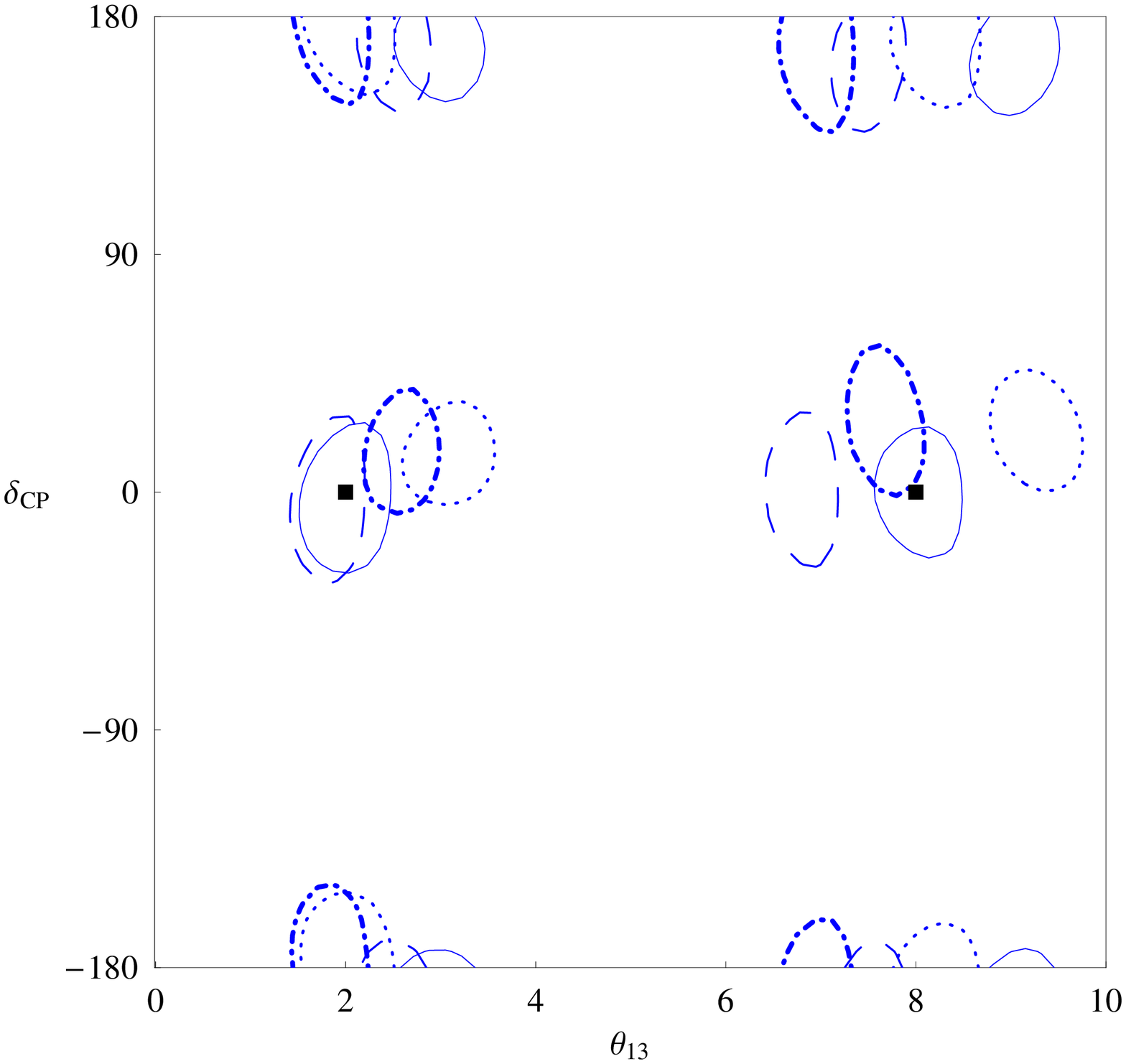} &
\hspace{-0.5cm} \epsfxsize8.25cm\epsffile{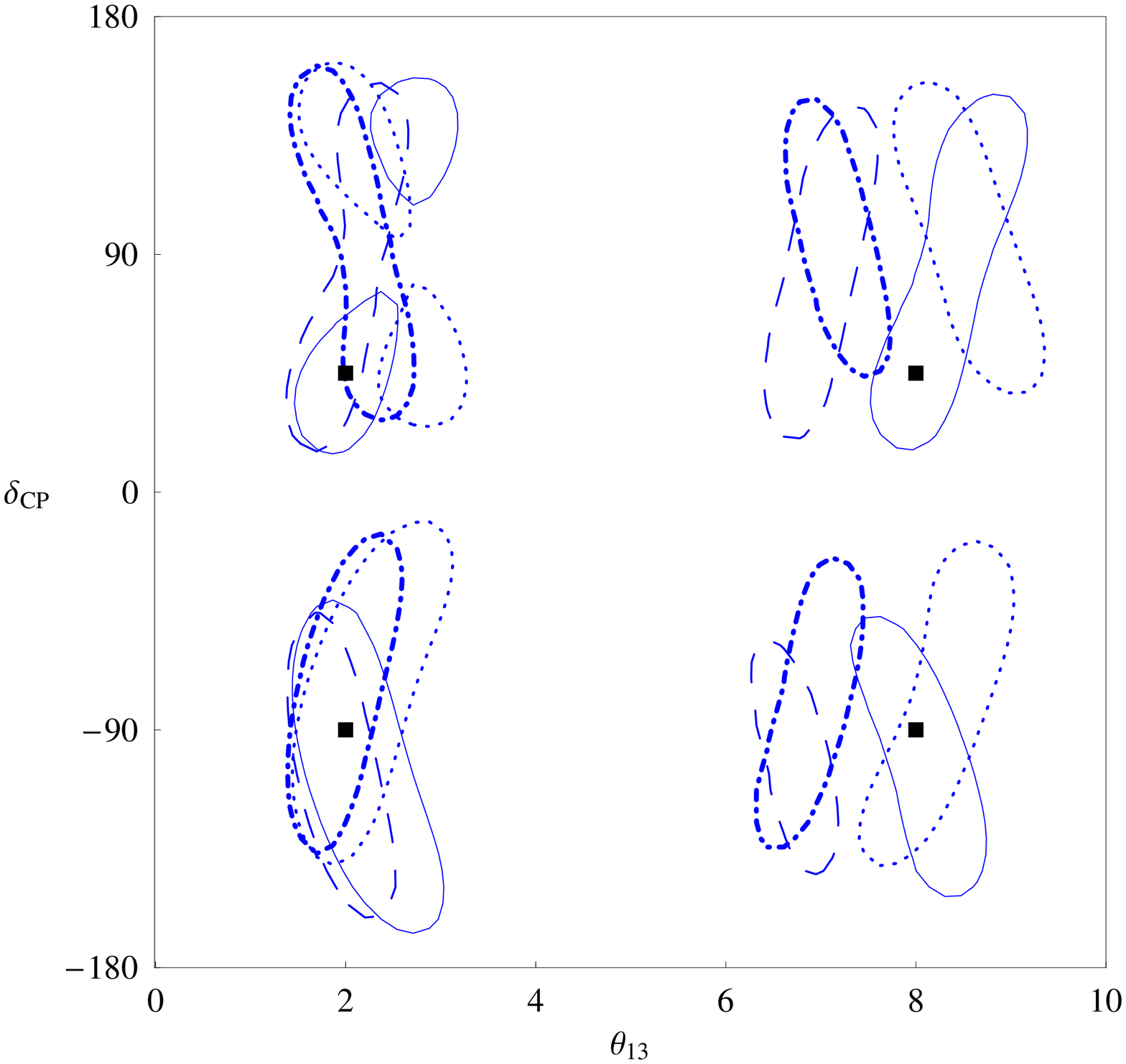} \\
\end{tabular}
\caption{\it 
$90$\% CL contours in the ($\theta_{13},\delta$) plane using the appearance channel
after a $10$ years run at the \BB with a $440$ kton detector located at $L = 130$ km, 
for two different values of $\theta_{13}$, $\bar \theta_{13} = 2^\circ,8^\circ$, 
and three values of $\delta$, $\bar\delta=0^\circ$ (left plot) and $\bar\delta=45^\circ,
-90^\circ$ (right plot). A $5$\% systematic error is assumed and backgrounds are 
computed as in ref.~\cite{Donini:2004hu}.
Continuous, dotted, dashed and dot-dashed lines stand for the intrinsic, sign, octant 
and mixed degeneracies, respectively.
}
\label{fig:appBB}
\end{center}
\end{figure}

The measurement of $(\theta_{13},\delta)$ at this facility has been already actively 
discussed in the literature \cite{Apollonio:2002en,allBB}. In particular, a complete 
analysis of the eightfold degeneracy was done in \cite{Donini:2004hu}. In Fig.~\ref{fig:appBB} 
we plot our results for three different CP phases, $\bar \delta = 0^\circ$ (left plot) 
and $\bar \delta = 45^\circ,- 90^\circ$ (right plot), and for two different mixing angles 
$\bar \theta_{13} = 2^\circ$ and $8^\circ$. The input $(\bar \theta_{13}, \bar \delta)$ 
value used in the fit is always shown as a filled black box. As in \cite{Donini:2004hu}, 
we use the following reference values for the atmospheric and solar parameters: 
$\Delta m^2_{atm} = \Delta m^2_{23} = 2.5 \times 10^{-3}$ eV$^2$, $\theta_{12} = 33^\circ$ and 
$\Delta m^2_{sol} = \Delta m^2_{12} = 8.2 \times 10^{-5}$ eV$^2$ \cite{globalfit}. 
The atmospheric mixing angle, 
$\theta_{23}$, has been fixed at $\theta_{23} = 40^\circ$, thus inducing the so-called
octant degeneracies\footnote{This will be the case if the T2K experiment observes non-maximal 
mixing in the atmospheric sector. Notice that a choice of $\theta_{23} = 50^\circ$ would
give similar results.}. 
The $90$\% CL contours for each of the degenerate solutions are depicted in the plot: 
continuous lines stand for $s_{atm}=\bar s_{atm},s_{oct}=\bar s_{oct}$ (the {\it true} 
solution and the {\it intrinsic} clone); dotted lines stand for $s_{atm}=-\bar s_{atm},
s_{oct}=\bar s_{oct}$ (the {\it sign} clone); dashed lines stand for $s_{atm}=\bar s_{atm},
s_{oct}=-\bar s_{oct}$ (the {\it octant} clone); dot-dashed lines stand for $s_{atm}=
-\bar s_{atm},s_{oct}=-\bar s_{oct}$ (the {\it mixed} clone). These plots are obtained 
assuming a $5$\% systematic error. Backgrounds have been computed as in \cite{Donini:2004hu}.

In Fig.~\ref{fig:appBB} it can be seen the dramatic impact that degeneracies have 
in the precision of the measure of $(\theta_{13},\delta)$: 
(1) the knowledge of $\theta_{13}$ is worsened by, roughly, a factor four (two) for 
    large\footnote{The shift in $\theta_{13}$ of the clones with respect to the true 
    solution is proportional to $\theta_{13}$; see \cite{Donini:2003vz} for the explicit 
    derivation.} (small) values of $\theta_{13}$, as four possible separate 
    solution-regions appear. The presence of degeneracies has a small 
    impact on the ultimate $\theta_{13}$ sensitivity; 
(2) the knowledge of $\delta$ is worsened in a significant way in presence of the clones, 
    almost spanning half of the parameter space for small values of $\theta_{13}$. 
These facts are well understood. From the appearance channel of a counting experiment, 
like the standard \BB, with a baseline of hundreds of km (i.e practically in vacuum) 
there are not enough independent informations to cancel any of the degeneracies. 
We can also rephrase this fact in the following ``statistical'' way: the clones have always 
the same $\chi^2$ of the true solution, making impossible any discrimination between true 
solution and the degeneracies.  

It has been claimed \cite{reactor} that the \nuebar disappearance channel at a 
reactor experiment can help Super-Beam experiments in solving part of the 
eightfold degeneracy. Indeed, the \nue disappearance probability does not 
depend on the CP violating phase $\delta$ and the atmospheric $\theta_{23}$ 
mixing angle. Thus, the $\theta_{13}$ measurement is not affected by $\theta_{13}-\delta$ 
correlations nor by the octant and mixed ambiguities. The $\nue \! \! \raw \nue$ matter 
oscillation probability, expanded at second order in the small parameters $\theta_{13}$ 
and $(\Delta m^2_{sol} L/E)$ reads \cite{Akhmedov:2004ny}:
\bea
P_\mp (\nue \raw \nue) & = & 1\ -\ \left( \frac{\Delta_{atm}}{B_\mp} \right)^2 
       \sin^2 2\theta_{13} \ \sin^2 \left(\frac{B_\mp L}{2} \right) \ - \ 
       \left(\frac{\Delta_{sol}}{A} \right)^2 \sin^2 2 \theta_{12} \ \sin^2 
       \left(\frac{A L}{2}\right ) \nn \\
\label{disnue}
\eea 
where $\Delta_{atm}=\Delta m^2_{atm}/ 2 E,\Delta_{sol}=\Delta m^2_{sol} / 2 E$ and 
$B_\mp=|A \mp \Delta_{atm}|$ with $\mp$ for neutrinos (antineutrinos), respectively.
The dependence on the sign of the atmospheric mass difference arises from the first 
non-trivial term of eq.~(\ref{disnue}) and from higher order terms, 
${\mathcal O}(\theta_{13}^2 
\times \Delta m^2_{12} L/E)$. As a consequence, the $s_{atm}$ dependence is relevant 
for large values of $\theta_{13}$, only. 

\begin{figure}[t]
\vspace{-0.5cm}
\begin{center}
\begin{tabular}{cc}
\hspace{-1.0cm} \epsfxsize8.25cm\epsffile{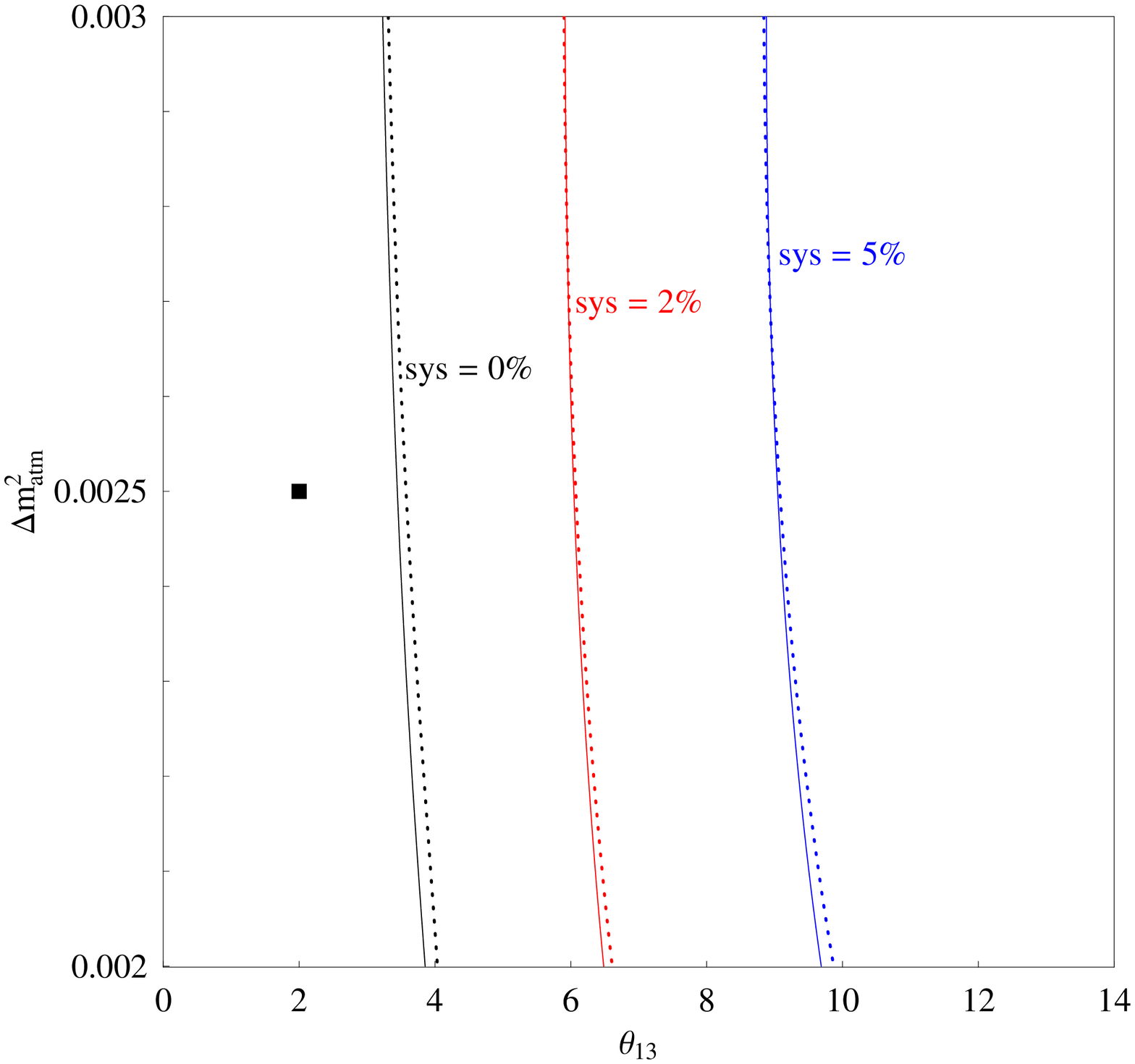} &
\hspace{-0.5cm} \epsfxsize8.25cm\epsffile{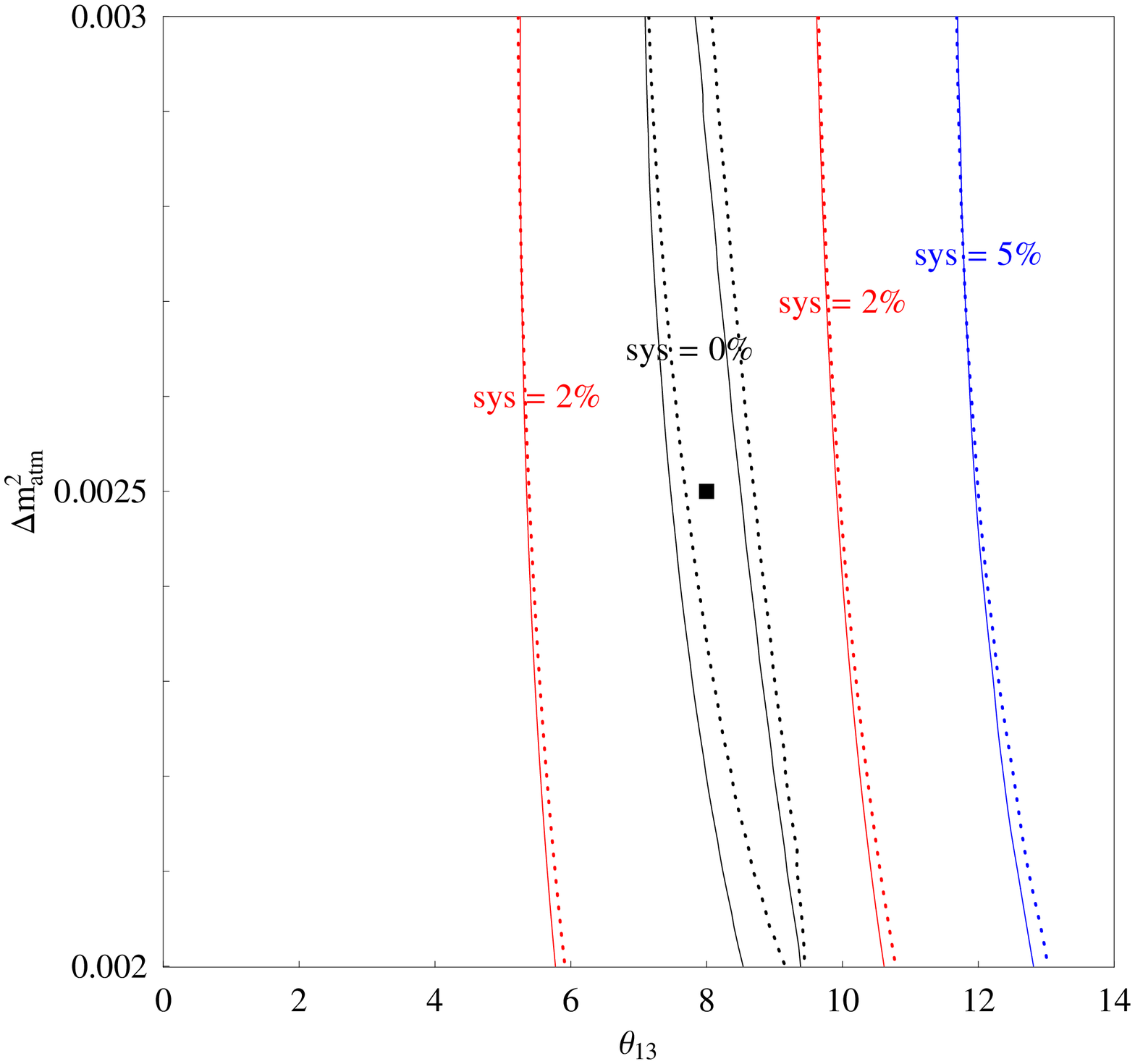} \\
\end{tabular}
\caption{\it 
$90$\% CL contours in the ($\theta_{13},\Delta m^2_{atm}$) plane using the disappearance 
channel after a $10$ years run at the \BB for two different values of $\theta_{13}$, 
$\bar\theta_{13}=2^\circ$ (left plot) and $\bar \theta_{13} = 8^\circ$ (right plot).
Three different values of the systematic errors have been considered: $0$\% (``theoretical''),
$2$\% (``pessimistic''), $5$\% (``optimistic''). Continuous lines stand for the true solution, 
dotted lines stand for the sign degeneracy. 
}
\label{fig:disBB}
\end{center}
\end{figure}

In Tab.~\ref{tab:betabeam} we summarize the relevant numbers for the \BB 
disappearance analysis. In the appearance analysis having a realistic description 
of backgrounds and efficiencies is of fundamental importance for providing the 
correct sensitivity on $(\theta_{13},\delta)$. Conversely, their relevance is 
much smaller in the disappearance measure, that is limited primarily by 
systematic errors. Lacking a complete realistic description of systematics 
we decided to present in Fig.~\ref{fig:disBB} the disappearance measure for three 
different systematic errors, namely $0$\% (``theoretical-unrealistic'' scenario), 
$2$\% (``optimistic'' scenario) and $5$\% (``pessimistic'' scenario). We decided to 
show the $0$\% systematic line, as we think it is ``theoretically'' important to have 
an idea of the ultimate reach of this experiment. The $2$\% and $5$\% lines will 
cover the optimistic and pessimistic feelings about future experimental improvements 
in understanding a Mton water detector.  

In Fig.~\ref{fig:disBB} the 90\% CL contours in the $(\theta_{13},\Delta m^2_{atm})$ plane 
using the \nue disappearance channel are shown for the input values $\bar \theta_{13} = 
2^\circ,8^\circ$ and $\Delta m^2_{atm}=2.5 \times 10^{-3}$ eV$^2$. The left plot 
($\bar \theta_{13} = 2^\circ$) represents, in practice, the $\theta_{13}$ sensitivity 
reach of the \BB disappearance channel.
If systematic errors cannot be controlled better than the $5$\% level, the \BB disappearance 
channel alone does not improve significantly the present bound on $\theta_{13}$.
The ``theoretical'' sensitivity (sys=$0$\%) is around $\theta_{13} = 4^\circ$, while if a 
systematics of $2\%$ is achieved the disappearance channel alone could test $\theta_{13}$ 
down to $6^\circ$. In Fig.~\ref{fig:disBB}(right) we show our results for a large value of 
$\theta_{13}$, $\bar \theta_{13}=8^\circ$. With a systematics of $2\%$ the mixing angle 
can be measured in the disappearance channel alone with an error of $\pm 2^\circ$. Again 
if the $5\%$ systematics is assumed no improvement on the present bound is obtained.

As it is clear from Fig.~\ref{fig:disBB}, the \nue disappearance channel is only slightly 
sensitive to the sign clone, as the full and dashed lines are almost superimposed for 
every value of $\theta_{13}$ and $\Delta m^2_{atm}$. The \nue disappearance channel is 
an almost ``clone-free'' environment for the $\beta$-Beam, as it is for reactor experiments. 
However, even in the case of an optimistic (but non-zero) 2\% systematic error, no 
improvement is obtained adding the disappearance channel informations to the results of 
Fig.~\ref{fig:appBB} for the appearance channel. The resulting 90\% CL contours practically 
coincide with the previous ones, and for this reason we do not consider to present them 
in a separate figure. The $\theta_{13}$ indetermination coming from the clone presence in 
the appearance channel is smaller than the disappearance error itself. Only considering 
an unrealistic 0\% systematics the disappearance channel starts to be useful to eliminate 
some of the clones.
%
%

Summarizing, the \BB has two available oscillation channels: the $\nue \!\! \raw \nue$ 
disappearance and the $\nue \!\! \raw \numu$ appearance (and their CP-conjugates). The 
appearance channel can measure $(\theta_{13},\delta)$, but being the considered \BB a 
pure counting experiment this measurement is severely affected by degeneracies. The 
disappearance channel does not provide any further useful informations once realistic 
systematic errors are taken into account. 
%
%
\section{Super-Beam Appearance and Disappearance Channels}
%
%
The considered \SB setup is a conventional neutrino beam based on the 4 MWatt SPL 2.2 GeV 
proton driver that has been proposed at CERN, described in ref.~\cite{Gomez-Cadenas:2001eu}.
The average neutrino energies of the $\nu_\mu,\bar \nu_\mu$ beams corresponding to this 
configuration are 0.27 GeV and 0.25 GeV, respectively. The possibility to measure 
$(\theta_{13},\delta)$ with a Super-Beam has been already widely discussed in the 
literature \cite{allSB}. A complete analysis of the eightfold degeneracy at this 
facility has been done in \cite{Donini:2004hu}. 

\begin{figure}[t]
\vspace{-0.5cm}
\begin{center}
\begin{tabular}{cc}
\hspace{-1.0cm} \epsfxsize8.25cm\epsffile{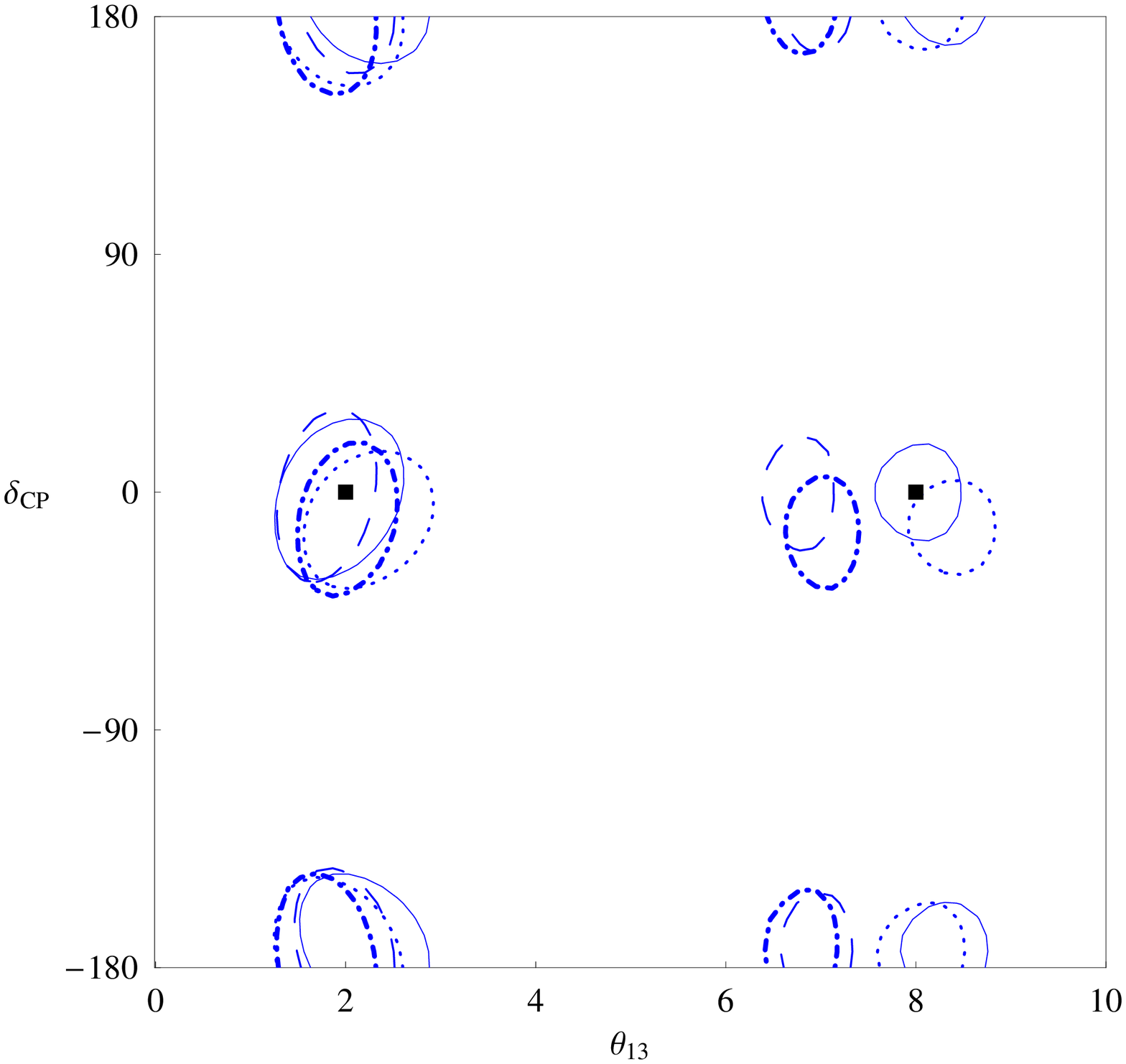} &
\hspace{-0.5cm} \epsfxsize8.25cm\epsffile{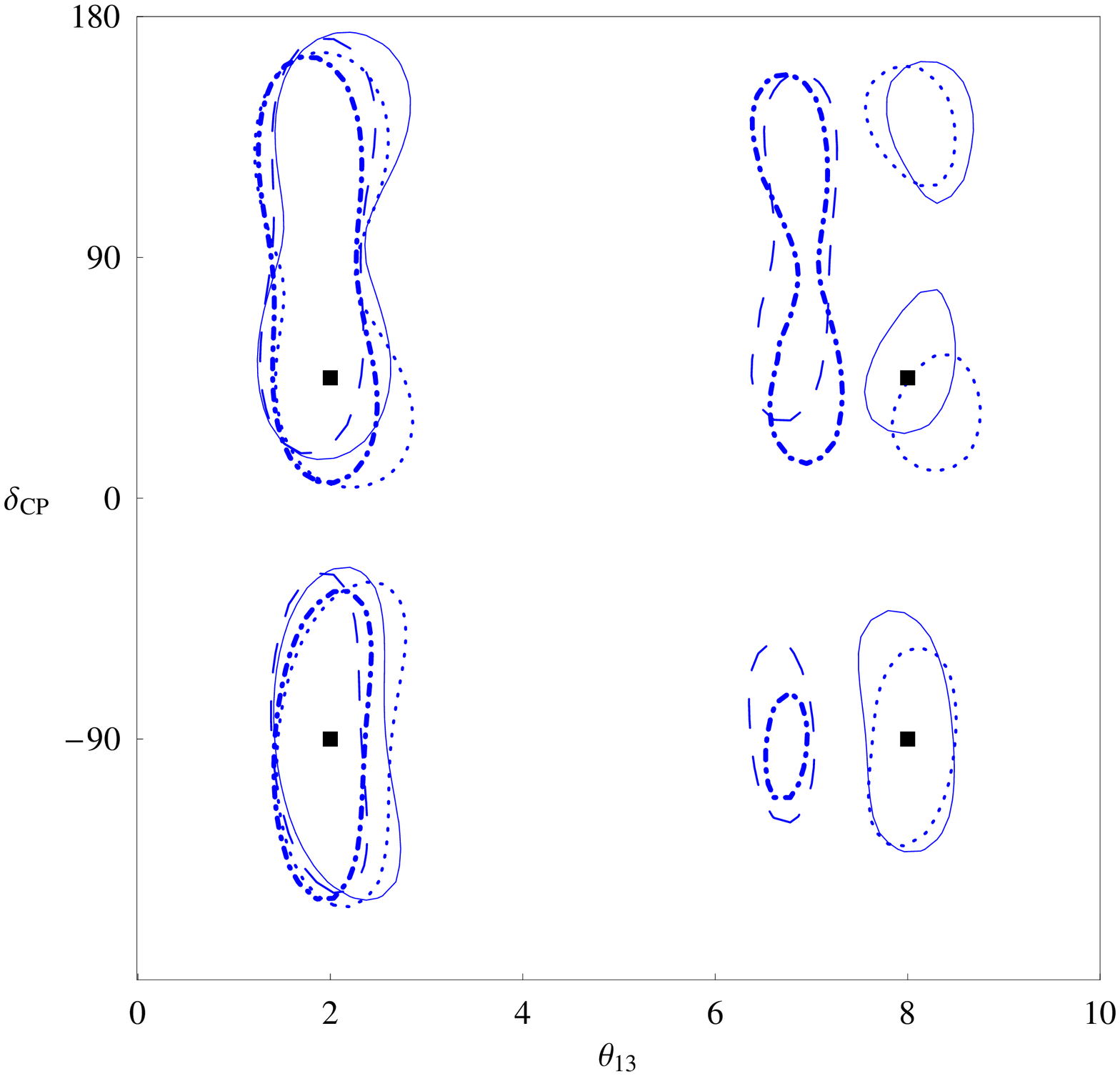} \\
\end{tabular}
\caption{\it 
$90$\% CL contours in the ($\theta_{13},\delta$) plane using the appearance channel
after a $2+8$ years run at the \SB with a $440$ kton detector located at $L = 130$ km, 
for two different values of $\theta_{13}$, $\bar \theta_{13} = 2^\circ,8^\circ$, 
and three values of $\delta$, $\bar \delta = 0^\circ$ (left plot) and $\bar \delta = 
45^\circ, -90^\circ$ (right plot). A $5$\% systematic error is assumed and backgrounds 
are computed as in ref.~\cite{Donini:2004hu}.
Continuous, dotted, dashed and dot-dashed lines stand for the intrinsic, sign, octant 
and mixed degeneracies, respectively.
}
\label{fig:appSB}
\end{center}
\end{figure}

In Fig.~\ref{fig:appSB} we plot our results for three different CP phases, $\bar\delta=
0^\circ$ (left plot) and $\bar \delta = 45^\circ, -90^\circ$ (right plot), and for two 
different mixing angles, $\bar \theta_{13} = 2^\circ$ and $8^\circ$. The input 
$(\bar\theta_{13},\bar\delta)$ value used in the fit is shown as a filled black box. 
As in the previous section we use the following reference values for the atmospheric and 
solar parameters: $\theta_{23}=40^\circ$, $\Delta m^2_{atm}=2.5 \times 10^{-3}$ eV$^2$, 
$\theta_{12}=33^\circ$ and $\Delta m^2_{sol}=8.2 \times 10^{-5}$ eV$^2$. The 
$90$\% CL contours for each of the degenerate solutions are depicted in the 
plot and explained in the caption. These plots are obtained assuming a $5$\% 
systematic error. Backgrounds have been computed~as~in~\cite{Donini:2004hu}.

As it appears from comparison of Fig. \ref{fig:appSB} with Fig. \ref{fig:appBB}, 
the ``figures of merit'' of a standard $\beta$-Beam and the SPL Super-Beam are 
very similar. Also the Super-Beam appearance channel is severely affected by 
proliferation of clones. The precision in measuring $(\theta_{13},\delta)$ is 
practically identical in the two cases. This is well explained by the comparable
statistics in the golden channel ($\nue \!\!\raw \numu$ vs $\numu \!\!\raw \nue$) 
and an almost equal $(L/E)$ ratio for the two experiments. 
For this reason there could be no real synergy between this two setups 
(i.e. they are not complementary) and the only effect in summing these two experiments, 
concerning the $(\theta_{13},\delta)$ measure, is to double the statistics. 
\begin{figure}[t!]
\vspace{-0.5cm}
\begin{center}
\begin{tabular}{cc}
\hspace{-1.0cm} \epsfxsize8.25cm\epsffile{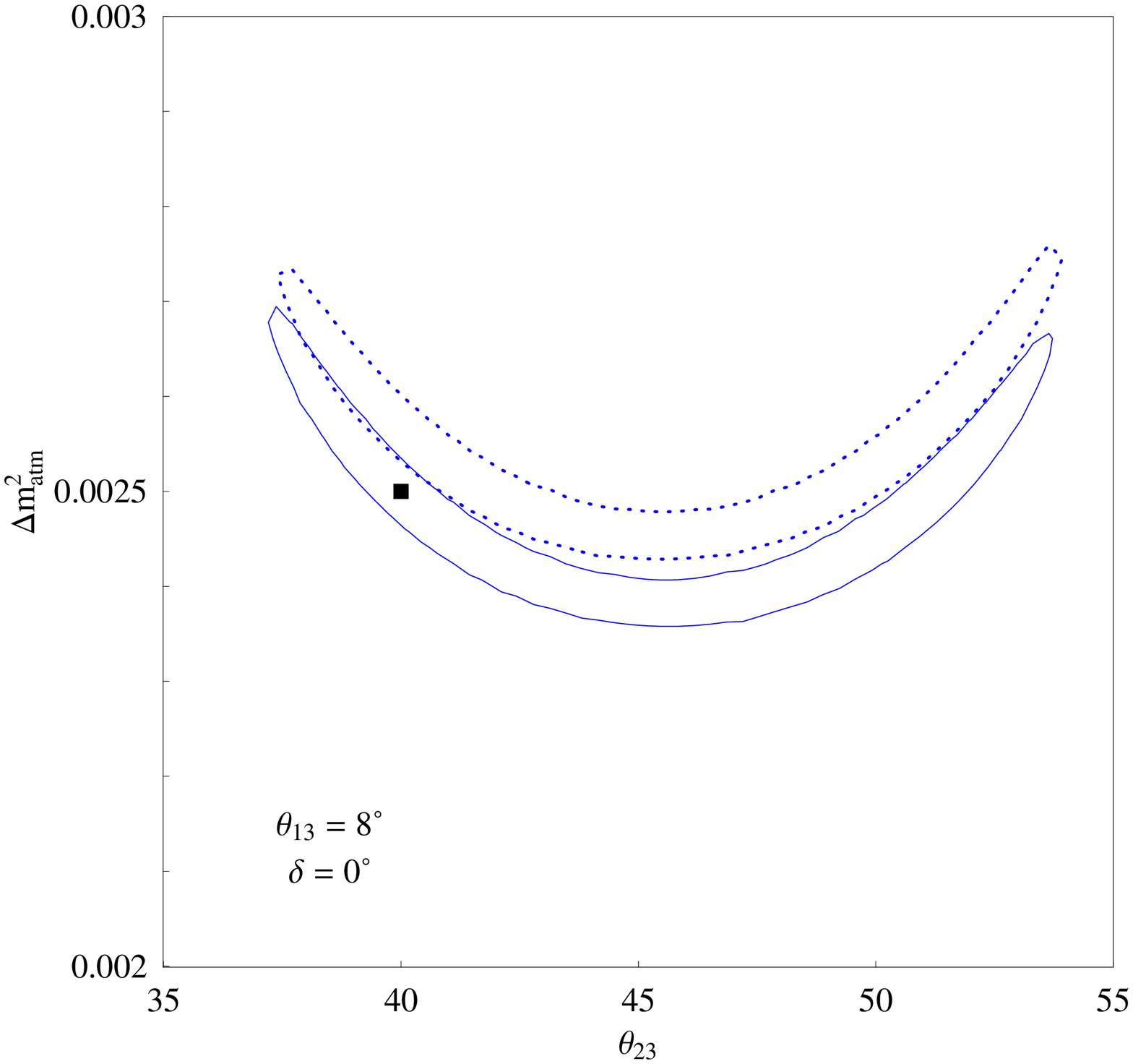} &
\hspace{-0.5cm} \epsfxsize8.25cm\epsffile{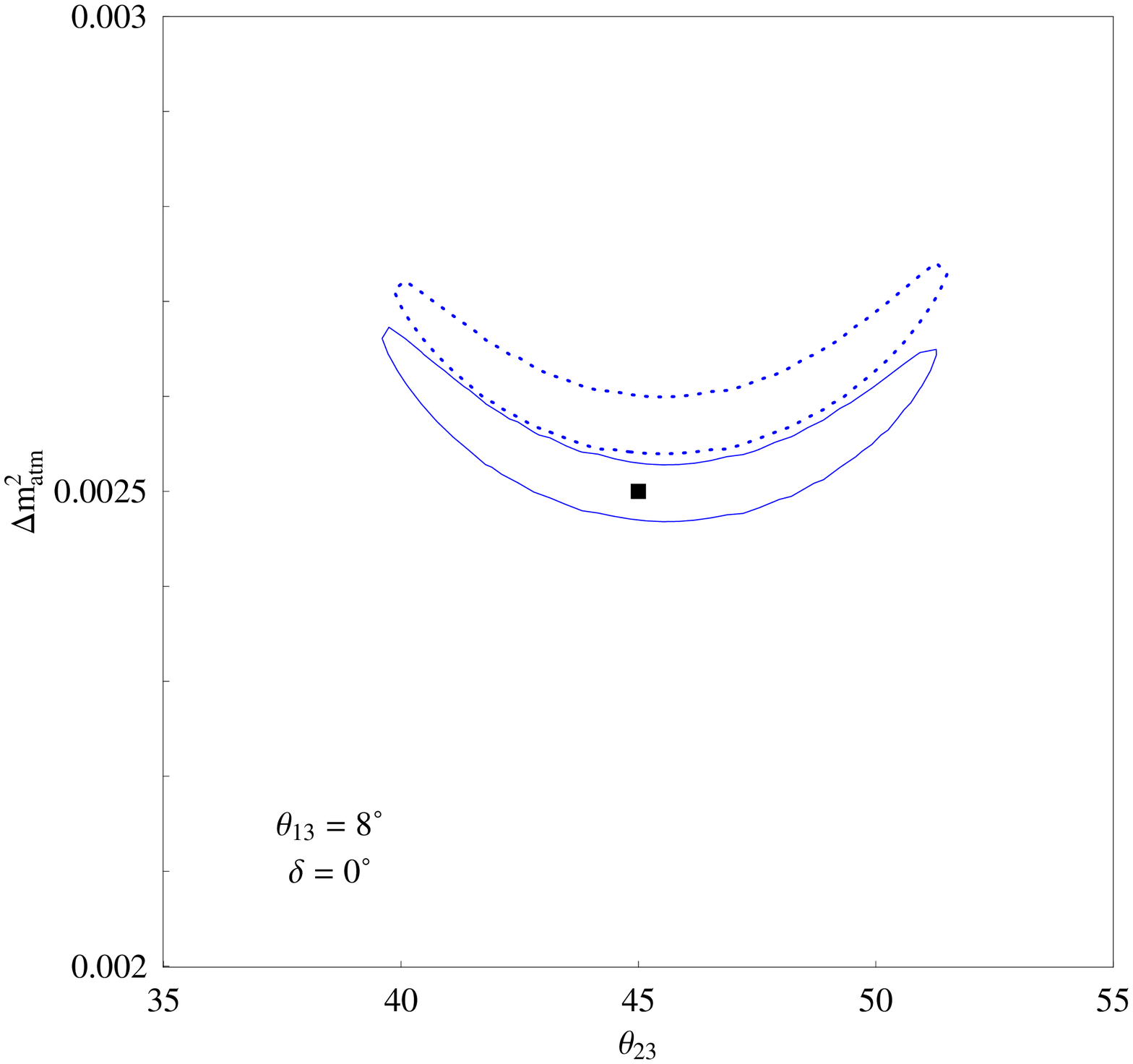} \\
\end{tabular}
\caption{\it 
$90$\% CL contours in the $(\theta_{23},\Delta m^2_{atm})$ plane using the disappearance channel 
after a $2+8$ years run at the $4$ MWatt SPL \SB for two different values of $\theta_{23}$, 
$\theta_{23} = 40^\circ$ (left plot) and $\theta_{23} = 45^\circ$ (right plot).
A systematic error of $2$\% is assumed. 
Continuous lines stand for $s_{atm} = \bar s_{atm}$; 
dotted lines stand for $s_{atm} = - \bar s_{atm}$.
}
\label{fig:disSB}
\end{center}
\end{figure}

Nevertheless, the great advantage of the \SB facility compared with the 
standard-$\gamma$ \BB one is the possibility to measure directly the atmospheric 
parameters using the $\numu$ disappearance channel \cite{Minakata:2004pg} reducing, 
in particular, the atmospheric mass difference error to less than 10\%. 
\begin{table}[hbtp]
\begin{center}
\begin{tabular}{|c|c|c|c|c|c|} \hline
 $N_{\mu^-}$ & No Osc. & $\theta_{13} = 8^\circ; +$ & $\theta_{13} = 8^\circ; -$ & 
                         $\theta_{13} = 2^\circ; +$ & $\theta_{13} = 2^\circ; -$\\ \hline \hline
$\delta=0^\circ$        & 24245 & 2016 & 2197 & 1987 & 2136 \\ \hline
$\delta=90^\circ$ &       & 2037 & 2175 & 1993 & 2131 \\ \hline
\hline
 $N_{\mu^+}$ & No Osc. & $\theta_{13} = 8^\circ; +$ & $\theta_{13} = 8^\circ; -$ & 
                         $\theta_{13} = 2^\circ; +$ & $\theta_{13} = 2^\circ; -$\\ \hline \hline
$\delta=0^\circ$        & 25467 & 1982 & 2178 & 1944 & 2095 \\ \hline
$\delta=90^\circ$ &       & 2009  & 2150 & 1951 & 2088 \\ \hline
\hline
\end{tabular}
\end{center}
\caption{\it 
Disappearance event rates for a $2+8$ years run at the $4$ MWatt SPL \SB with a $440$ kton 
detector at $L=130$ km, for different values of $\theta_{13},\delta$ and of the sign of 
the atmospheric mass difference, $s_{atm}$. Appearance event rates have been presented 
in ref.~\cite{Donini:2004hu}.
}
\label{tab:superbeam}
\end{table}

In Tab.~\ref{tab:superbeam} we summarize the relevant numbers used in the \SB 
disappearance analysis. In Fig. \ref{fig:disSB} we show the measure of $(\theta_{23},
\Delta m^2_{atm})$ at the SPL \SB with a 2\% systematic error, in the case (left) of 
non-maximal atmospheric mixing, $\theta_{23} = 40^\circ$, and (right) maximal atmospheric 
mixing, $\theta_{23} = 45^\circ$. In both cases, the input value for the atmospheric mass 
difference has been fixed to $\Delta m^2_{atm}=2.5 \times 10^{-3}$ eV$^2$. The continuous 
contour represents the fit to the right choice of the sign of the atmospheric mass 
difference (i.e. $s_{atm} = \bar s_{atm}$) whereas the dotted contour represents the fit to 
the wrong choice of $s_{atm}$ (i.e. $s_{atm}=-\bar s_{atm}$). Since we plot the results 
in the full $\theta_{23} \in [35^\circ-55^\circ]$ parameter space, the octant and mixed 
clones in the left plot are automatically taken into account and do not appear as separate
regions. 
Notice, however, that being the contours for 
$\theta_{23} \leq 45^\circ$ and $\theta_{23} \geq 45^\circ$ slightly different for 
$\theta_{13} \neq 0^\circ$, if we were to plot the contours in the $(\sin^2 2 \theta_{23}, 
\Delta m^2_{atm})$ plane a fourfold degeneracy would be manifest. In the right plot only 
the sign clone is present, being $\theta_{23} = 45^\circ$. 
Two comments are in order: first, the sign ambiguity implies that the errors on the 
atmospheric mass difference are roughly doubled with respect to what expected in the 
absence of degeneracies; second, the left plot is significantly worse than the right 
plot. If $\theta_{23}$ is not maximal, the errors on the atmospheric parameters 
$(\theta_{23},\Delta m^2_{atm})$ can be much larger than expected. 

The presence of degeneracies in the \numu disappearance channel 
can be easily understood looking at the the $\numu \! \! \raw \numu$ 
vacuum oscillation probability expanded to the second order in the small 
parameters $\theta_{13}$ and $(\Delta m^2_{sol} L/E)$, \cite{Akhmedov:2004ny}:
\bea
P(\numu \raw \numu) & = & 1 - 
    (\sin^2 2 \theta_{23} - s^2_{23} \sin^2 2 \theta_{13} \cos 2 \theta_{23} ) \
      \sin^2\left(\frac{\Delta_{atm} L}{2} \right) \nn \\
   & - &       \left(\frac{\Delta_{sol} L}{2} \right) \ 
       \left[s^2_{12} \sin^2 2\theta_{23} + \tilde{J} s^2_{23} \cos\delta \right] \
       \sin \left(\Delta_{atm} L \right) \nn \\
   & - &  \left(\frac{\Delta_{sol} L}{2} \right)^2 
          \left[c^4_{23} \ \sin^2 2\theta_{12} + 
               s^2_{12} \ \sin^2 2\theta_{23} \ \cos \left(\Delta_{atm} L \right) \right] 
\label{disnumu}
\eea 
where $\tilde{J} = \cos\theta_{13}\sin 2\theta_{12}\sin 2\theta_{13}\sin 2\theta_{23}$. 
The first non-trivial term is the dominant (atmospheric) contribution. It does not depend 
on the solar parameters and it reduces to the usual two-family approximation when 
$\theta_{13}=0^\circ$. The last term is the subleading solar contribution, suppressed by 
two powers of the solar mass difference. This term is independent (in this approximation) 
from $\theta_{13}$. Eventually, the term in the second line is the interference between the 
atmospheric and the solar contributions: it is small but not negligible compared to the 
first term, being suppressed by only one power of the solar mass difference. This term 
encodes both a $\theta_{13}$-dependence (through the $\tilde J$ coefficient) and a small 
CP-conserving $\delta$-dependence (suppressed by one power of $\Delta_{sol}$ and one power 
of $\theta_{13}$). Changing the sign of the atmospheric mass difference makes the interference 
term change sign, also, mimicking an increase of $\Delta m^2_{atm}$. Notice, finally, that 
the three non-trivial terms in eq.~(\ref{disnumu}) are not symmetric for $\theta_{23} \raw 
\pi/2 -\theta_{23}$. However, the non-symmetric dependence on $\theta_{23}$ is suppressed 
by at least two powers of $\theta_{13}$, $\Delta_{sol}$ or their combination, making the 
asymmetry extremely small. 
\begin{figure}[t!]
\vspace{-0.5cm}
\begin{center}
\begin{tabular}{cc}
\hspace{-1.0cm} \epsfxsize8.25cm\epsffile{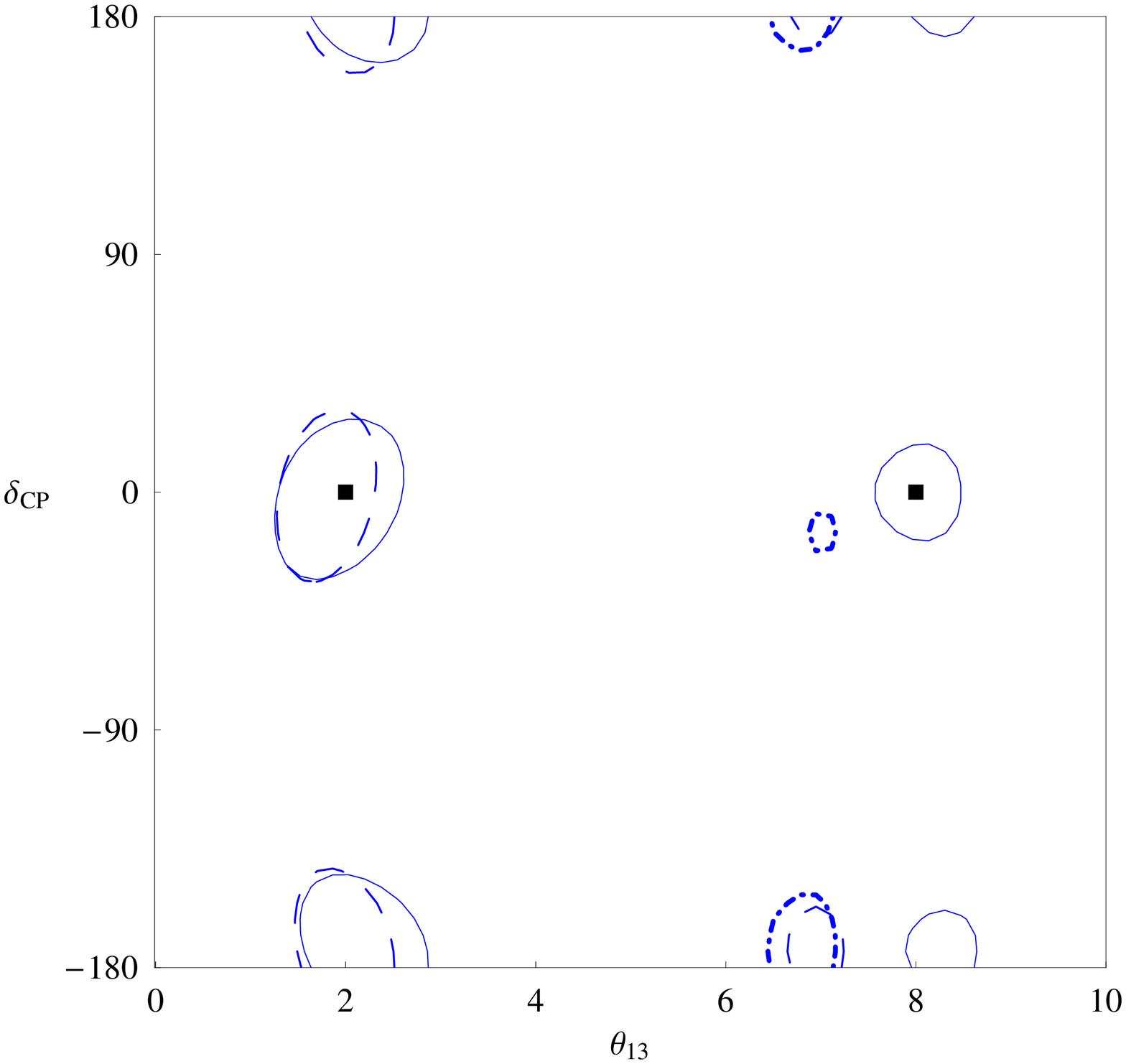} &
\hspace{-0.5cm} \epsfxsize8.25cm\epsffile{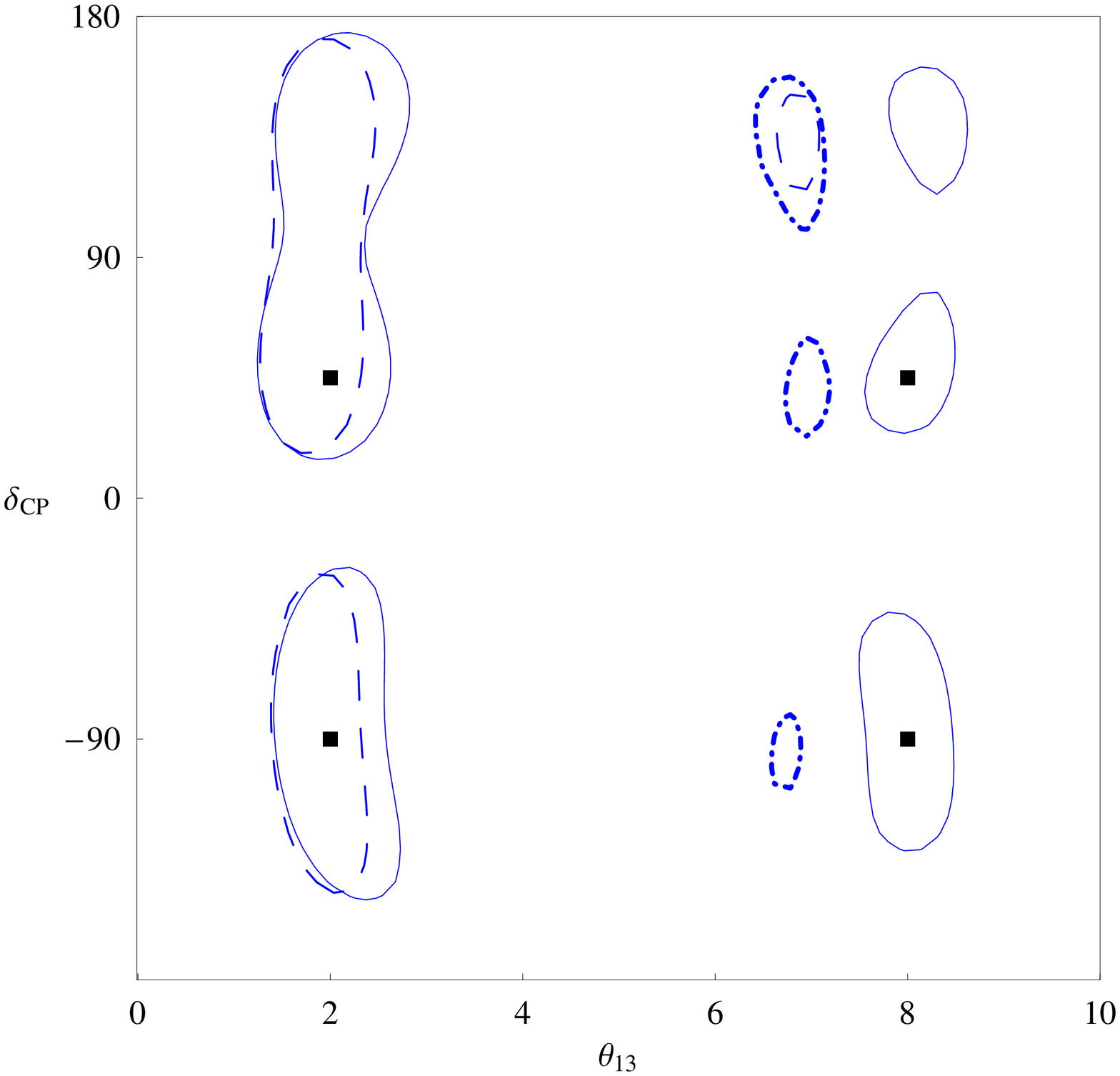} \\
\end{tabular}
\caption{\it 
$90$\% CL contours in the ($\theta_{13},\delta$) plane using the appearance and the 
disappearance channels after a $2+8$ years run at the \SB with a $440$ kton detector 
located at $L = 130$ km, for two different values of $\theta_{13}$, $\bar\theta_{13}=
2^\circ,8^\circ$, and three values of $\delta$, $\bar \delta = 0^\circ$ (left plot) and 
$\bar \delta = 45^\circ, -90^\circ$ (right plot). A $5$\% ($2$\%) systematic error is 
assumed for the appearance (disappearance) channel, and backgrounds are computed as in 
ref.~\cite{Donini:2004hu}.
Continuous, dotted, dashed and dot-dashed lines stand for the intrinsic, sign, octant 
and mixed degeneracies, respectively.
}
\label{fig:totSB}
\end{center}
\end{figure}

In Fig.~\ref{fig:totSB} we present the simultaneous measurement of $(\theta_{13},\delta)$ 
using both the appearance (with a 5\% systematic error) and the disappearance 
(with a 2\% systematic error) channels at the Super-Beam. 
As it can be noticed, contrary to the $\beta$-Beam case, the disappearance channel 
in the \SB fit introduces significant changes. Notably enough, the sign clone has disappeared 
in any case considered. This is not a surprise as these fits are performed at a fixed 
$\Delta m^2_{atm}$: since in the disappearance channel the sign clone manifests itself 
at a larger value of $\Delta m^2_{atm}$ (see Fig.~\ref{fig:disSB}), in the combination with 
the appearance channel the tension between the two suffices to remove the unwanted clone 
in the $(\theta_{13},\delta)$ plane. Notice, moreover, that in some cases the octant clone 
is considerably reduced or even solved, due to the octant-asymmetric contributions in the 
disappearance probability, eq.~(\ref{disnumu}). Nonetheless, this does not mean that thanks 
to the combination of the appearance and the disappearance channels we are indeed able to 
measure the sign of the atmospheric mass difference, $s_{atm}$. The mixed clones are generally 
still present for large values of $\theta_{13}$, thus preventing us from measuring $s_{atm}$, 
if the $\theta_{23}$-octant is not known at the time the experiment takes place.

\begin{figure}[t!]
\vspace{-0.5cm}
\begin{center}
\begin{tabular}{cc}
\hspace{-1.0cm} \epsfxsize8.25cm\epsffile{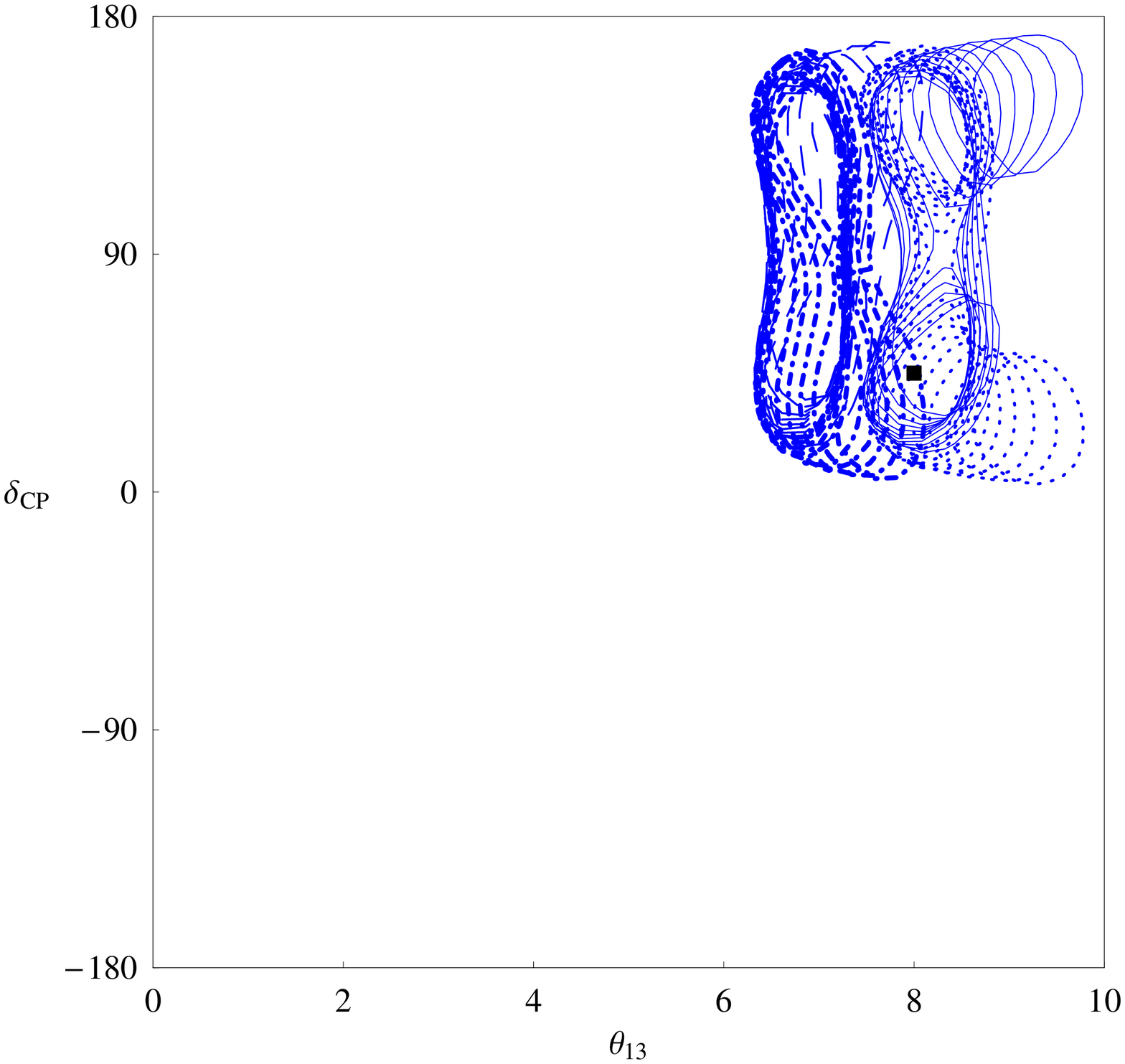} &
\hspace{-0.5cm} \epsfxsize8.25cm\epsffile{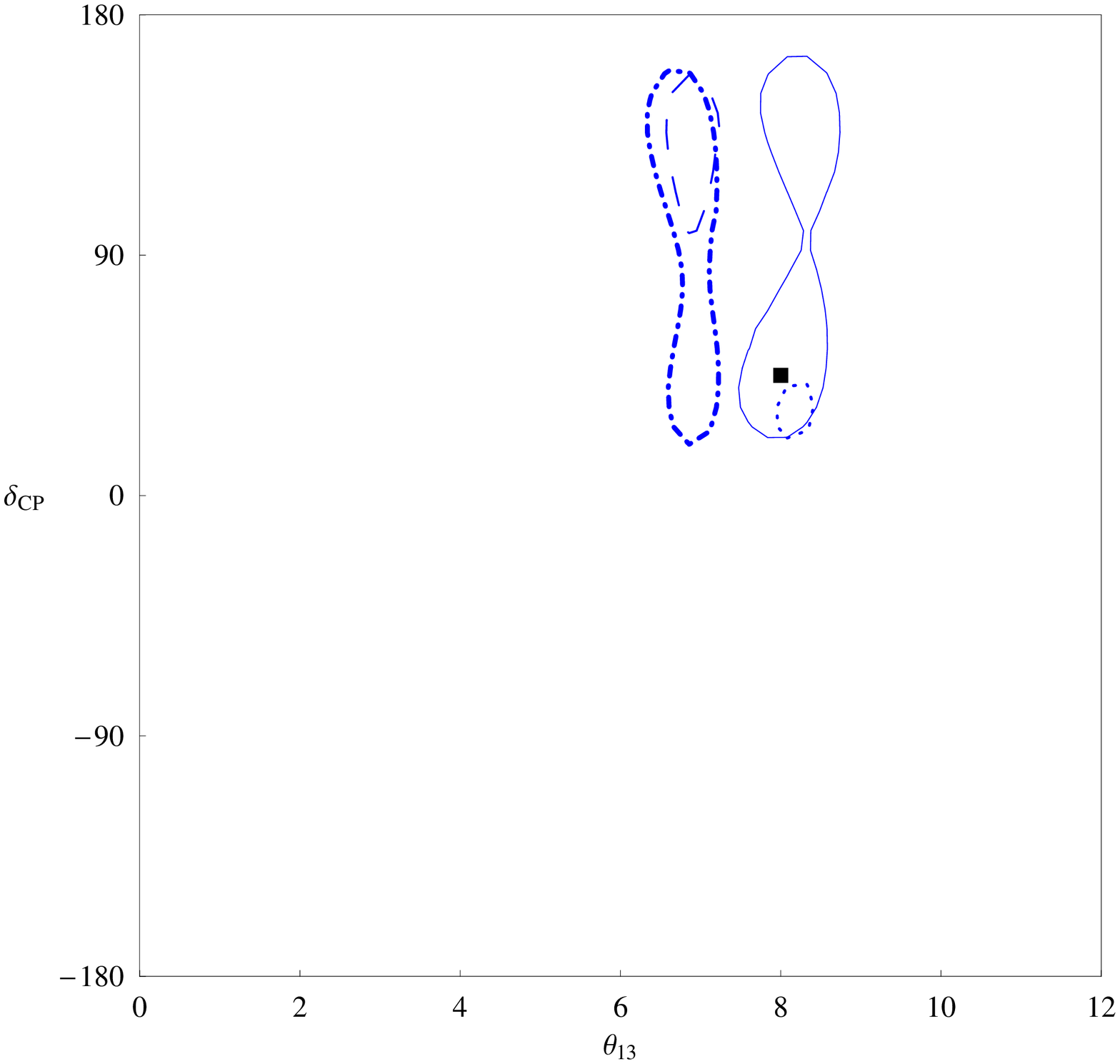} \\
\end{tabular}
\caption{\it 
Projection on the $(\theta_{13},\delta)$ plane of the $90$\% CL three-dimensional contours
for the \SB appearance (left) and appearance plus disappearance (right) channels. 
The fit has been performed in ($\theta_{13},\delta, \Delta m^2_{atm}$) 
with input values $\bar \theta_{13} = 8^\circ$, $\bar \delta = 45^\circ$ 
and $\Delta m^2_{atm} = 2.5 \times 10^{-3}$ eV$^2$, for $\theta_{23} = 40^\circ$. 
}
\label{fig:totSB3D}
\end{center}
\end{figure}

It is clear that these results should be confirmed by a complete multi-dimensional analysis 
that is actually underway \cite{meloni}. As a first step, in Fig.~\ref{fig:totSB3D}(left) 
we show the projection on the $(\theta_{13},\delta)$ plane of the \SB appearance 
three-dimensional fit in the parameters $(\theta_{13},\delta,\Delta m^2_{atm})$ for the 
following input values: $\bar \theta_{13} = 8^\circ$, $\bar \delta = 45^\circ$ and 
$\Delta m^2_{atm} = 2.5 \times 10^{-3}$ eV$^2$. We let $\Delta m^2_{atm}$ varying freely 
in the range $\Delta m^2_{atm} \in [2.0 - 3.0] \times 10^{-3}$ eV$^2$, obtained from 
Fig.~\ref{fig:disSB}. In Fig.~\ref{fig:totSB3D}(right) we show the same three-dimensional 
fit, adding this time both the \SB appearance and disappearance channels. It can be seen 
that in a complete three-parameters analysis small remnants of the sign clones are still 
present, contrary to the case of the two-parameters analysis of Fig.~\ref{fig:totSB}. It is, 
however, still true that the degeneracy structure gets strongly reduced\footnote{It must 
be added that the \BB option could give similar results (see \cite{meloni}) to that 
presented in Fig.~\ref{fig:totSB3D}, once complemented by a \SB disappearance channel 
such as SPL or T2K-I \cite{T2K}.}. 
%
%

%
\section{CP Discovery Potential}
%

\begin{figure}[t!]
\vspace{-0.5cm}
\begin{center}
\begin{tabular}{cc}
\hspace{-1.0cm} \epsfxsize8.5cm\epsffile{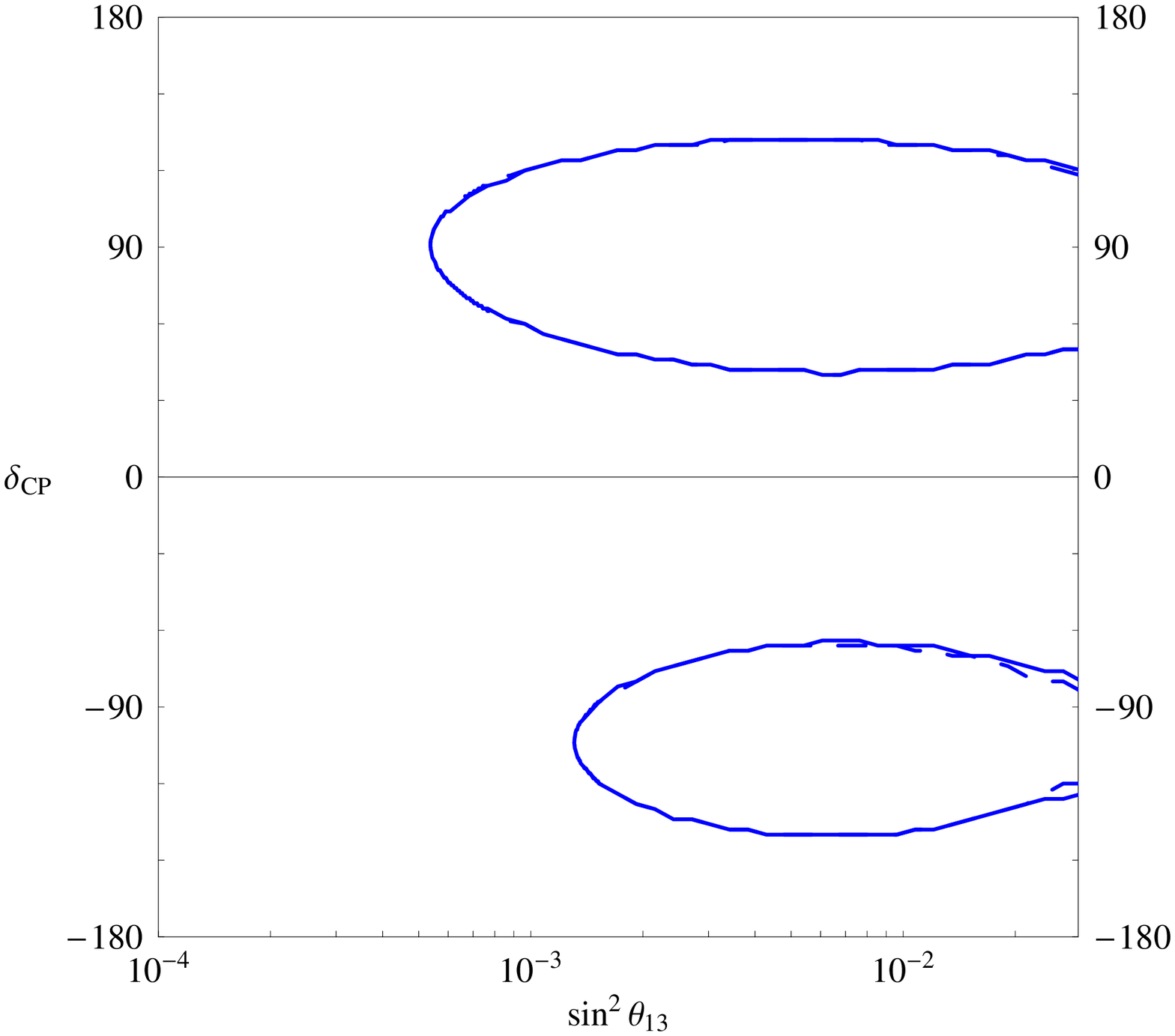} &
\hspace{-0.5cm} \epsfxsize8.5cm\epsffile{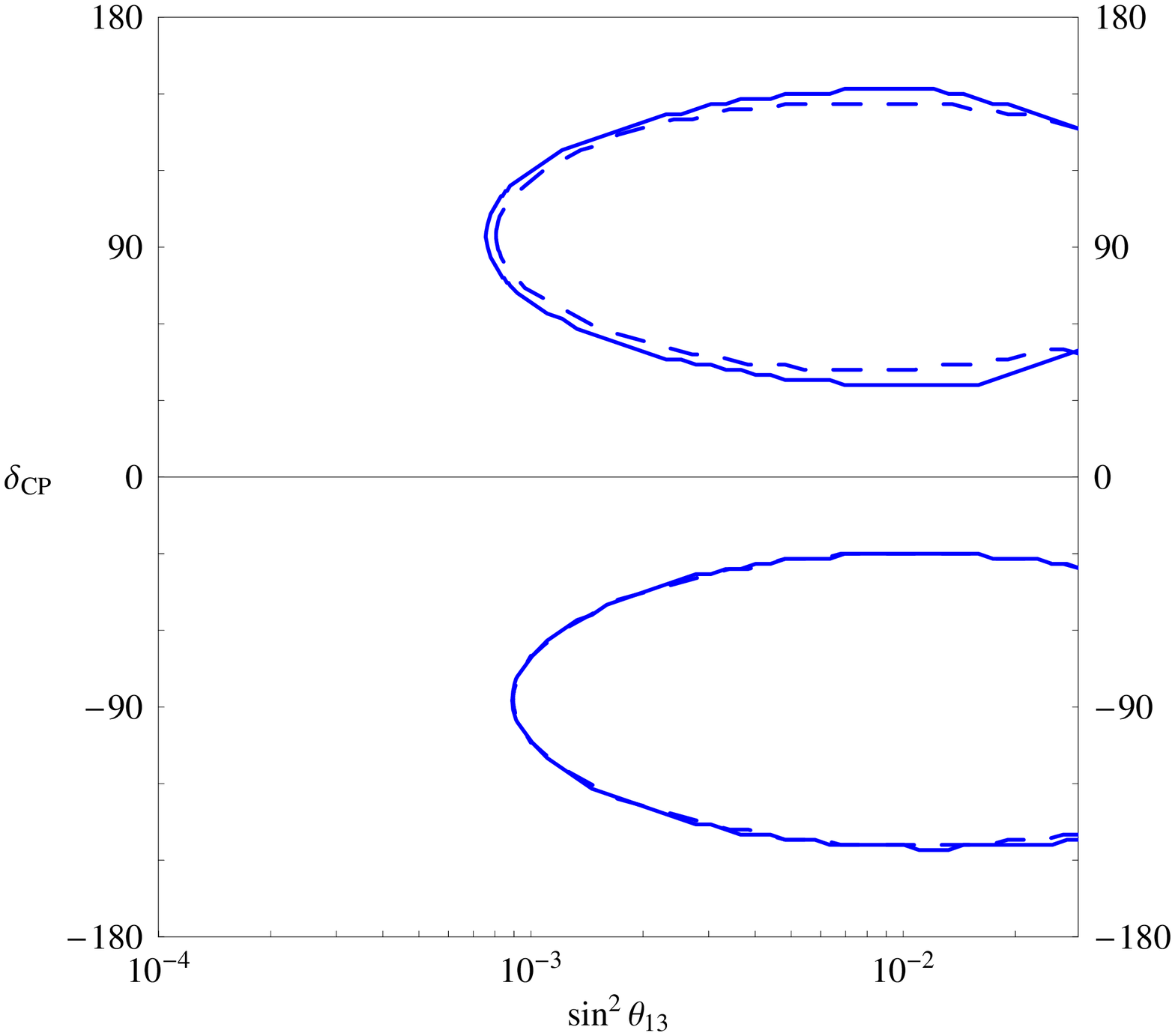} \\
\end{tabular}
\caption{\it 
$3\sigma$ CP discovery potential in the $(\theta_{13},\delta)$ plane for the considered 
\BB (left) and \SB (right). Dashed lines stand for appearance channel only, whereas solid 
lines stand for the combination of the appearance and the disappearance channels at the 
same facility. A $2$\% systematic error in the disappearance channel and a $5$\% systematic 
error in the appearance channel have been considered for both facilities.
}
\label{fig:sensitivity}
\end{center}
\end{figure}

Eventually, in Fig.~\ref{fig:sensitivity} we present the CP discovery potential to 
$(\theta_{13},\delta)$ using the appearance channel only or the combination of the 
appearance and the disappearance channels, for the $\beta$-Beam (left) and the \SB 
(right). The 3$\sigma$ contours have been computed as follows: at a fixed $\bar 
\theta_{13}$, we look for the smallest (largest) value of $|\bar \delta|$ for which 
the two-parameters 3$\sigma$ contours of any of the degenerate solutions (true, sign, 
octant and mixed) do not touch $\delta = 0^\circ$ nor $\delta = 180^\circ$. Notice that, 
although the input $\bar \theta_{13}$ value is fixed, the clones can touch $\delta = 
0^\circ, 180^\circ$ at $\theta_{13} \neq \bar \theta_{13}$, also\footnote{This is not 
the case of Fig.~11 in ref.~\cite{Donini:2004hu}, where the excluded region in $\delta$ 
at fixed $\bar \theta_{13}$ in the absence of a CP-violating signal at 90\% CL is 
presented. In practice, in that figure we compare $N_\pm (\bar \theta_{13},\delta)$ 
with $N_\pm (\bar \theta_{13}, 0^\circ)$, thus obtaining a one-parameter sensitivity 
plot in $\delta$ only.}. The outcome of this procedure is finally plotted, representing 
the region in the $(\theta_{13},\delta)$-parameter space for which a CP-violating 
signal is observed at 3$\sigma$. The novelty of this plot, with respect to Fig.~6 of 
\cite{Bouchez:2003fy} and to Fig.~13 of \cite{Burguet-Castell:2003vv}, is that we 
fully take into account the impact of the parameter degeneracies to derive the 
CP-violation discovery power of these facilities. Moreover, we present the results 
for the whole allowed range in $\delta$, $\delta \in \left[ -180^\circ, 180^\circ 
\right]$. This is particularly appropriate, since only an approximate symmetry is 
observed for $|\delta| \ge \pi/2$ and $|\delta| \le \pi/2$ and no symmetry at all 
between positive and negative $\delta$. For both facilities, we have applied a 2\% 
systematic error on the disappearance channel (i.e., $\nu_\mu \to \nu_\mu$ and $\nu_e 
\to \nu_e$) and a 5\% systematic error on the appearance channel (i.e., $\nu_e\to 
\nu_\mu$ and $\nu_\mu\to \nu_e$).

First of all, notice that the \SB CP discovery potential in $\theta_{13}$ is symmetric 
in $\delta$ (for both sectors, we observe an ultimate sensitivity of $\sin^2 
\theta_{13} \simeq 6-8 \times 10^{-4}$). This is not the case for the $\beta$-Beam, 
where for positive $\delta$ the facility outperforms the \SB and for negative 
$\delta$ is outperformed by it. We know, however, that this asymmetric behaviour of 
the \BB for positive and negative $\delta$ is a statistical  mirage caused by the low 
background in the appearance antineutrino sample and the high background in the 
appearance neutrino one (see \cite{Donini:2004hu}). A proper statistical treatment 
should be performed, following \cite{Feldman:1997qc}, to get rid of this asymmetry 
in the small $\sin^2 \theta_{13}$ case: the treatment, however, is extremely time 
consuming and we do not consider meaningful applying it here. Regarding the discovery 
potential to $\delta$, notice how for large $\theta_{13}$ the \SB is generally performing 
better than the $\beta$-Beam, in particular for negative $\delta$. This can be understood 
comparing the right plots in Fig.~\ref{fig:appBB} and Fig.~\ref{fig:totSB}: for $\bar 
\theta_{13} = 8^\circ, \bar \delta = -90^\circ$ we can see that the 90\% CL contours 
for the \SB are much smaller than for the $\beta$-Beam. For positive $\delta$ the 
difference is not so relevant, since although the \SB contours for $\bar \theta_{13} = 
8^\circ, \bar \delta = 45^\circ$ are certainly smaller than the \BB ones, it can be 
seen that the spread in $\delta$ is similar. 

Eventually, we stress that for the \SB a small improvement in the discovery potential 
in $\delta$ is achieved combining appearance and disappearance channels. On the contrary, 
practically no effect is observed in the \BB case.

%
\section{Conclusions}
%
%
Summarizing, in this letter we have tried to understand the impact of a disappearance 
measurement on the $(\theta_{13},\delta$) eightfold degeneracy for two specific 
facilities, the 4 MWatt SPL \SB and the standard low-$\gamma$ \BB proposed at CERN. 
We presented a complete analysis of degenerations in the \nue and \numu disappearance 
channels: the \nue disappearance is affected by a twofold degeneracy, since the \nue 
probability depends on $s_{atm}$ only, eq.~(\ref{disnue}); the \numu disappearance is 
affected by a fourfold degeneracy, depending on both $s_{atm}$ and $s_{oct}$, eq.~(\ref{disnumu}). 

The standard low-$\gamma$ $\beta$-Beam setup looks somewhat limited, when compared with 
facilities with many channels to exploit, such as the Neutrino Factory: indeed, the {\em 
golden} $\nue \! \raw \numu$ appearance channel is severely affected by degeneracies 
(being the neutrino energy too low to use energy resolution techniques) and the combination 
with the $\nue \! \raw \nue$ disappearance, potentially of interest, is in practice useless 
once a realistic systematic error is taken into account. The \BB idea should be certainly 
pursued further, using for example higher $\gamma$ options. For neutrino energies around 
1 GeV is, in fact, possible to take advantage of energy resolution \cite{Itow:2001ee}
and, for energies higher than 4-5 GeV, the silver channel $\nu_e \to \nu_\tau$ becomes 
available. In both cases, the different informations can be used to reduce the parameter 
space degeneracies and solve some of the clones, consequently improving our knowledge on 
$\theta_{13}$ and $\delta$ beyond the \SB reach.

The SPL \SB appearance channel, $\nu_\mu\to \nu_e$, is also severely affected by degeneracies 
(being a counting experiment, as the $\beta$-Beam). However, in this case the complementarity 
between the appearance and disappearance channels, $\nu_\mu\to \nu_e$ and $\nu_\mu\to \nu_\mu$, 
can be fully exploited even when a realistic systematic error is taken into account. In particular, 
the sign ambiguity can be strongly reduced (Fig.~\ref{fig:totSB3D}), a consequence of the fact 
that the disappearance sign clone is located at a different $\Delta m^2_{atm}$ for different 
choices of $s_{atm}$, Fig.~\ref{fig:disSB}. Notice, finally, that the $\nu_\mu$ disappearance 
channel is interesting on its own for a precise measurement of the atmospheric oscillation 
parameters, $(\theta_{23},\Delta m^2_{atm})$, Fig.~\ref{fig:disSB}. 
   
It is clear that, in the case where only one of the two facilities were to be built at 
CERN, the 4 MWatt SPL \SB would represent a more interesting choice than the standard
low-$\gamma$ $\beta$-Beam to study the leptonic mixing matrix. This is because the two 
experiments have a similar discovery potential to $(\theta_{13},\delta)$ and \SB can add 
useful informations from the disappearance \numu channel. Anyway, it is now evident that 
the \SB race against T2K is going to be lost. So one should concentrate on different 
beam technologies, \BB being one of the options. We believe that a higher-$\gamma$ \BB 
should be considered instead of the standard one, in such a way that, adding spectral 
information, a measure of $(\theta_{13},\delta)$ preciser than at the \SB could be 
obtained.  
%
\section*{Acknowledgments}
%
We would like to thank B.~Gavela, J.~Gomez-Cadenas, P.~Hernandez, D.~Meloni 
and P.~Migliozzi for useful discussions. We acknowledge the financial support 
of MCYT through project FPA2003-04597 and of the European Community-Research 
Infrastructure Activity under FP6 "Structuring the European Research Area" 
programme (CARE, contract number RII3-CT-2003-506395).
%

%
%

%
%
\end{document}